# MU-MIMO Grouping For Real-time Applications

Hannaneh B. Pasandi, Tamer Nadeem
Hadi Amirpour

## ABSTRACT

Over the last decade, the bandwidth expansion and MU-MIMO spectral efficiency have promised to increase data throughput by allowing concurrent communication between one Access Point and multiple users. However, we are still a long way from enjoying such MU-MIMO MAC protocol improvements for bandwidth hungry applications such as video streaming in practical WiFi network settings due to heterogeneous channel conditions and devices, unreliable transmissions, and lack of useful feedback exchange among the lower and upper layers' requirements. This paper introduces MuViS, a novel dual-phase optimization framework that proposes a Quality of Experience (QoE) aware MU-MIMO optimization for multi-user video streaming over IEEE 802.11ac. MuViS first employs reinforcement learning to optimize the MU-MIMO user group and mode selection for users based on their PHY/MAC layer characteristics. The video bitrate is then optimized based on the user's mode (Multi-User (MU) or Single-User (SU)). We present our design and its evaluation on smartphones and laptops using 802.11ac WiFi. Our experimental results in various indoor environments and configurations show a scalable framework that can support a large number of users with streaming at high video rates and satisfying QoE requirements.

## 1 INTRODUCTION

Recent research has shown that multi-gigabit WiFi technologies like 802.11ac and upcoming 802.11ax or LTE [24] can meet the strict latency, reliability, and quality of experience (QoE) requirements of bandwidth-hungry applications like augmented and virtual reality (AR/VR), the next frontier of mobile devices [3, 12, 14, 15, 18], and video streaming [25, 31].

Multi-User Multiple-Input and Multiple-Output (MU-MIMO) in 802.11ac is a technique that improves spectral efficiency by allowing concurrent transmission between a single access point (AP) and multiple clients. Because it allows concurrent downlink transmissions to multiple clients, MU-MIMO has a significant potential benefit.

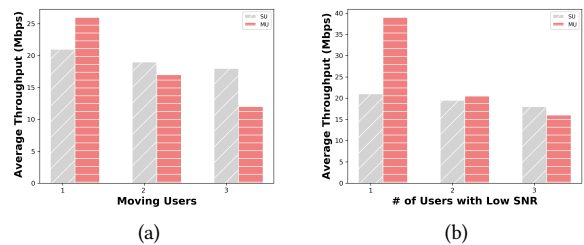

Figure 1: SU versus MU gains in average throughput for varying numbers of (a) moving users, (b) users with low SNR.

However, if the wrong users[1] are grouped together in MU-MIMO transmission, the grouping protocol may introduce higher delays and lower throughput (high packet losses). While recent studies [21, 23] experimentally demonstrated that it is not uncommon for MU-MIMO to under-perform Single-User-MIMO (SU-MIMO) in various scenarios. To demonstrate this, we conducted two experiments to compare the average throughput of three users in SU vs. MU mode for two different factors when users were a) mobile and b) had low SNR values. The experiment was carried out on 5G band with 20Mhz, and the downlink was saturated by UDP traffic using the devices and AP listed in Table 1. As shown in Figure 1, as the number of users with mobility or low SNR values increases (as indicated by the indices in the x-axis), the MU-MIMO average throughput begins to degrade. Such observation implies that incorrect grouping can result in incorrect MCS selection in the PHY layer, lower video bitrate, and poor QoE in the application layer [11]. These findings are consistent with [23], which shows that the beamforming SNR in MU-MIMO can be significantly lower than the SNR in SU-MIMO.

In this paper, we focus on multi-user video streaming applications over 802.11ac in a practical network environment. We begin with an in-depth analysis to understand the factors that may sabotage MU-MIMO grouping performance. We then describe how MU-MIMO grouping can have an impact on the upper layer performance. Later, we propose a *dual-phase* optimization approach for MU-MIMO grouping in a multi-user application scenario. While such optimization could be applied to other multi-user applications(e.g., mu-VR), we specifically focus on multi-user adaptive video streaming applications. We leave the impact of the device type itself for the future research.

---

[1]In this work, the terms user, receiver, and client are used interchangeably.

Video streaming techniques rely on Adaptive Bitrate (ABR) schemes to adjust the video playback for a diverse set of user devices and network conditions. Each video is partitioned into segments, where each segment includes a few seconds long playback. Each segment is then encoded in a number of different encoding rates (henceforth referred to as the video rates). When the user plays a video, the video player can download each segment at a video rate that is appropriate for its connection, therefore switching rates in response to changes in the available bandwidth. A number of rate adaptation schemes [27, 32] have been proposed to improve the user's QoE based on network throughput and/or buffer status that is locally observed at the user side. However, these techniques are mostly designed for a single-user setting. Despite a plethora of work on improving single user QoE, when multiple users are competing for the network resources, it has been shown that these schemes may result in video bitrate oscillation, network resource underutilization, or QoE unfairness among users [1, 13]. In mobile and wireless networks, a user adjusts the video quality based on the measured throughput that depends on their available downlink transmission rate. Since the transmission rate for each user is determined by wireless resource allocation schemes in the physical layer, an application-agnostic and/or unfair resource allocation in the MAC layer inevitably results in poor QoE. Some cross-layer approaches for multi-user adaptive video streaming [30] have been proposed to ensure fairness among DASH users, which jointly optimize the physical layer transmission rate for each user with QoE/bandwidth fairness objectives.

To solve the aforementioned performance degradation in both MU-MIMO MAC grouping and QoE in the application layer over WiFi, we take a cross-layer approach, MuViS which is *a dual-phase* framework for multi-user video streaming over IEEE 802.11ac. MuViS aims for MU-MIMO group selection in the MAC layer, which not only provides higher speeds (~433 to 1733 Mbps in practice), but also supports simultaneous transmission to multiple users at the same time. However, if the WiFi Access Point (AP) selects the incorrect users to group in a MU-MIMO transmission, it may introduce high delays and low throughput. Factors affecting MU-MIMO user grouping can consequently sabotage the performance of multi-user applications over 802.11ac/ax. By taking these factors into account, we first propose a Reinforcement Learning(RL)-based MU-MIMO group and mode optimization. Then, we propose a QoE optimization of ABR video streaming over WiFi using Lyapunov optimization technique [5]. As shown in [6], the channel of a user can be significantly different at a few centimeters apart in an indoor environment due to the rich multipath propagation. A user moving at walking speed may result in a significant change in the channel. Therefore, the goal of MuViS is to take into account these external factors (e.g., user mobility and device chipset characteristics) that typically are not considered in conventional methods when designing QoE optimization algorithms over WiFi.

Based on our experiments, MuViS can improve the QoE of multiple users even if their channel quality is unstable due to the undesirable external factors. The MuViS optimization framework across the MAC and application layers is briefly described in the following two phases:

**Phase I (§4.2.1)** We propose an RL-based MU-MIMO grouping and mode selection that takes into account not only the conventional factors that impact MU-MIMO grouping such as SNR, but also external factors such as user mobility (the motion of individual users can be tracked by reading their explicit Channel State Information (CSI) using a signal model such as [33]). By considering these factors as the input to the RL agent, MuViS learns how to optimize the MU-MIMO grouping and mode selection.

**Phase II (§??)** The underlying streaming algorithm adjusts the transmission and video rates in Phase II based on the desired QoE metrics (segment losses, buffer underflows, and video rate switches) specified by network operators or receivers. In doing this, MuViS formulates the QoE optimization problem for MU-MIMO aware wireless video streaming as a utility maximization problem using Lyapunov technique.

**Contributions.** Our core technical contributions could be summarized in the followings:

- MuViS is a novel dual-phase optimization framework that first uses reinforcement learning to learn the best set of MU-MIMO MAC grouping and mode selection and adapts when tested in unknown environments.
- We present an architecture and a practical testbed implementation which allow evaluating MuViS algorithm across different realistic indoor environments for variety of scenarios. MuViS is implemented and evaluated on commercial WiFi AP, COTS mobile devices, and laptops. When compared to the state-of-the-art, MuViS achieves a better performance in terms of QoE across different scenarios [10, 34].
- To the best of our knowledge, this work presents the first research proposal that takes a cross-layer approach to design online algorithms for QoE optimization of ABR video streaming over wireless by grouping the users based on their MU-MIMO MAC and PHY characteristics and their QoE requirements in the application layer.

## 2 BACKGROUND

This section provides a brief review of IEEE 802.11 ac enhancements, the beamforming procedure introduced by IEEE 802.11ac standard, the legacy user selection, and Reinforcement Learning (RL).

### 2.1 IEEE MU-MIMO MAC

**IEEE 802.11 ac vs. IEEE 802.11n** Compared to the previous IEEE 802.11n, IEEE 802.11ac (current WiFi) mainly introduces physical layer (PHY) enhancement. It operates on 5GHz, and supports denser modulation (up to 256-QAM). The use of MIMO is enhanced by increasing the maximum number of supported spatial streams from four to eight, and adding support for multi-user MIMO (MU-MIMO) together with a standardized approach to beamforming.

**MU-MIMO and Beamforming** The AP uses precoding technology to implement beamforming to generate strong signals in the respective direction to each user but weak signals in other directions to ensure good wireless coverage while mitigating interference caused by other users.

In particular, for each MU-MIMO transmission in 11ac the AP follows a sounding protocol by transmitting a Null Data Packet (NDP) announcement frame to gain channel access. This is followed by an NDP frame as in SU beamforming, so that the first user in the group can respond with a compressed beamforming action frame containing the measured CSI in compressed form. The AP then polls additional users in the group sequentially, each using a new beamforming report poll frame, to collect their respective *compressed beamforming reports*. The compressed beamforming action frame also includes an extra MU Exclusive Beamforming report on per subcarrier delta SNR along with an average SNR. Note that the sounding procedure introduces millisecond-level overhead to data transmissions [2], and the length of sounding overhead is dependent on the number of users in the MU-MIMO group, enabling frequent channel sounding leads to a non-trivial degradation in network throughput [4]. To reduce the impact of the sounding overhead, a practical network typically limits the sounding period on the order of a tenth of a second rather than enabling sounding in every transmission time-slots.

**MU-MIMO Grouping Optimization** An 802.11ac AP decides upon a set of users to transmit data concurrently through a user selection algorithm that precedes the MU-MIMO sounding and beamforming. User selection algorithm is not specified by the 802.11ac standard and it is AP vendor's implementation specific. Each user within an MU-MIMO group can operate with independent PHY rate, identified by a rate adaptation algorithm.

### 2.2 Reinforcement Learning

Reinforcement learning is a machine learning technique where the agent interacts with a time-variant *environment* that can be modeled as a Markov Decision Process (MDP), a Partially Observable MDP (POMDP), a game, etc. The core components of the RL technique are **environment**, **reward** ($r$), **possible set of actions** ($\mathcal{A}$) and **states** ($\mathcal{S}$). The state is the perception of the environment by the agent and is defined based on the sensory information of the agent. The agent selects an *action* from the given *state* and receives a *reward*. Actions are the agent's methods that allow it to interact and change its environment, and thus transfer between states. The *policy* $\pi$ prescribes actions to be taken in a given state. We can then value a given state $s$ and a policy $\pi$ in terms of expected future rewards.

Q-learning is one of the most popular off-policy algorithms in RL. It is also regarded as temporal difference learning that learns an *action-value* function to find a Q-value for each state-action pair. Q-learning agent learns its optimal policy by exploring and exploiting the environment. At each time instant $t$, the agent observes the current state $s_t$ and chooses a proper available action $a_t$ from this state to maximize the cumulative reward in time instant $t + 1$. More formally, the Q-value of $(s_t, a_t)$ from the policy $\pi$ which is denoted as $Q^\pi(s_t, a_t)$ is the sum of discounted reward received at time $t + 1$ when action $a_t$ is taken in state $s_t$, and it follows the optimal policy $\pi^*$, thereafter. The Q-values are updated using the following rule known as one of the Bellman equation forms:

$$Q(s_t, a_t) \leftarrow Q(s_t, a_t) + \alpha[r_{t+1} + \gamma \max_{a_{t+1}} Q(s_{t+1}, a_{t+1}) - Q(s_t, a_t)] \quad (1)$$

where $\alpha$ is the learning rate and $\gamma$ is the discount factor. Intuitively, the above equation adjusts the long-term or delayed rewards for a given state, action, and future state $s_{t+1}$ by weighting the previous Q-value estimate, the reward received and the best possible long-term reward obtained in the future state. The Q-values estimate can be adjusted for any delayed reward desired. For instance, to seek short-term rewards exclusively, we can set $\gamma = 0$, while to estimate the history of all rewards it would be $\gamma = 1$.

## 3 TRANSMISSION LINK AND NETWORK MODEL

We consider a downlink orthogonal frequency-division multiplexing (OFDM) network with one single-hop AP with a number of antenna denoted as $N_t$, a wireless network with $N$ receivers that each has a number of antenna denoted as $N_r$. Denote $S_g$ as the number of spatial streams of each receiver that is participate in MU-MIMO transmission. For precoding, we adopt zero-forcing technique [28] that has been widely adapted in 802.11ac technologies [17].

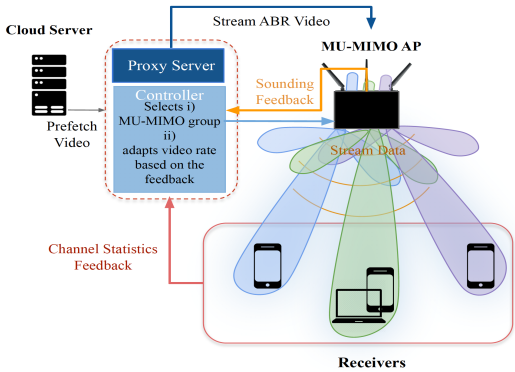

Figure 2: MuViS Framework Overview.

Precoding technique cancels the interference among spatial streams. The zero-forcing technique require a precise CSI which can be obtained from sounding period. Since frequent sounding overhead adds extra overhead [4] in a practical setting, the AP typically enables the sounding on the order of 1/10 of a second [19, 21].
seconds.

## 4 SYSTEM DESIGN, IMPLEMENTATION AND RESULTS: DUAL-PHASE OPTIMIZATION

In this section, we motivate why MU-MIMO grouping is a significant element to consider for a desirable video streaming QoE over WiFi. We then describe MuViS framework in more details and experimentally demonstrate the performance of the proposed DRL-based MU-MIMO grouping in Phase I and explain how the factors in this layer can impact the video streaming performance in the application layer.

### 4.1 Overview

As depicted in Figure 2, the architecture has four main components: a proxy server, WiFi AP, an AP controller, and receiver-side software. The proxy server interfaces with existing ABR services and locally caches the video content that is later transmitted from a commercial off-the-shelf AP. The video rate at the proxy server and the transmission rate at the AP are controlled by the MuViS framework. The receiver-side software is a lightweight application that does not require any modifications to the hardware or operating system. MuViS first performs Phase I which has two stages. In stage (i), it excludes the mobile users. In stage (ii) it learns to group the users using RL. Following the grouping decision in Phase I, MuViS then maximizes QoE while meeting constraints on the three QoE factors including lost segments, buffer underflows, and video rate switches as described in detail later in the text.

### 4.2 Phase I: MU-MIMO Grouping Optimization

---
**Algorithm 1** MuViS-I Algorithm
---
1: Initialize MU-MIMO group set $g$
2: Initialize $Q(s, a)$ arbitrary
3: **for** each episode **do**
4:     Select $a$ from policy derived from $Q$
5:     Take action $a$, calculate $r$ (throughput), observe $s'$
6:     $Q(s, a) \leftarrow Q(s, a) + \alpha[r + \gamma\, max(a)Q(s', a') - Q(s, a)]$
7:     $s \leftarrow s'$
8:     $g \leftarrow$ the combination that yields highest $r$
9: **end for**
---

Literature reports on a variety of factors that impact the performance of MU-MIMO grouping and mode selection in indoor and outdoor environments. Such factors include inter-user interference due to the lack of orthogonality between the channels of users [23], device type [21, 26], user mobility [21], and low SNR [21, 23]. However, pahse of optimization is mainly based on [7] and the ongoing work in this line of research.

**Why user mobility is important and how to track it?** As discussed in [21], most of the research on MU-MIMO grouping [20, 23, 29] assumes the obtained Channel State Information (CSI) readings from the limited sounding period is reliable and propose to group those users together that minimize an inter-client interference which is derived from CSI. In order to reduce the overhead of CSI feedback, the sounding interval in an 802.11ac network is on the order of one tenth of a second [19, 21]. Therefore, if a user moves at a walking speed between two consecutive sounding period, the last derived CSI is no longer valid in part of the interval between two sounding periods [22]. This unreliable CSI that is used for MU-MIMO beamforming leads to higher Packet Error Rate (PER) and throughput degradation when the moving users are part of the MU-MIMO group [20, 21]. To overcome this challenge, we perform two stages. In stage (i), we explicitly track each user's CSI and of movement is detected, we exclude the user from the group.

Motivated by the necessity to consider such factors in MU-MIMO grouping in the practical network setting, in stage (ii), we design MuViS-I. The underpinning technique in MuViS-I algorithm is Reinforcement Learning(RL). We explain our RL-based MU-MIMO grouping in the following.

*4.2.1 MU-MIMO Grouping as a RL problem.* We have formulated the MU-MIMO grouping as a reinforcement learning problem as following.

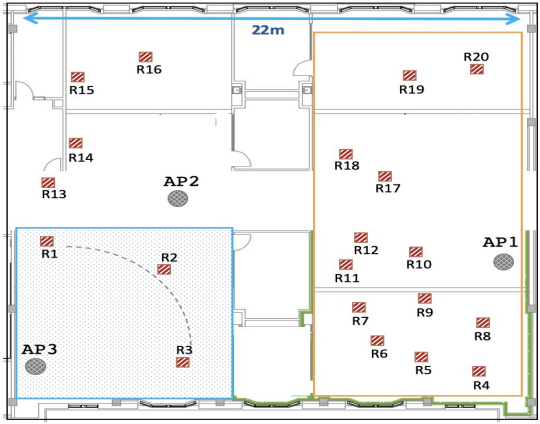

**Figure 3: Experimental floorplan. Locations of APs and receivers are marked.**

| List of Devices | Type | Count |
|---|---|---|
| Router | AP | 4 |
| Adaptor | Laptop | 3 |
| Adaptor (2x2 Stream) | Laptop | 3 |
| Linux | Laptop | 4 |
| Raspberry Pi 4 | Pi 4 | 3 |

**Table 1: List of devices in experiments.**

**Agent** is the AP that decides on MU-MIMO grouping and mode selection.

**State** is described as the vector of:
[state of the MU-MIMO groups, the number of the users and the number of spatial streams they support, user mobility status (0 if static, 1 otherwise)]

**Action** is to choose one of the possible combinations of groups and modes (SU or MU) that a user can be part of.

**Reward** is the downlink throughput.

These techniques are not directly applicable to MU-MIMO MAC downlink streaming, as they typically do not consider multiple receiver.

In recent times, another trend in video transmission is the learning-based methods [8, 9]. For example, Zwei [9] scheme, a self-play RL algorithm for video transmission tasks that aims to update the policy by utilizing the actual requirement. In another work, authors developed an ABR algorithm that utilizes DRL to select next chunks' bitrate [16]. We leave to explore this research direction for our future work.